\documentclass{aa}  
 \usepackage{graphicx}
 \usepackage{amssymb}
\usepackage{natbib}
\usepackage{booktabs}
\usepackage{multirow}
\usepackage{pdflscape}

\usepackage{color}
\newcommand{\kumiko}{\color{blue}}
\definecolor{color_fabrice}{rgb}{0.9,0.,0.}

\offprints{\email{kotera@iap.fr,  fabrice.mottez@obspm.fr }}

\title{Do asteroids evaporate near pulsars?\\ Induction heating by pulsar waves revisited}
\titlerunning{Induction heating by pulsar waves revisited}
\authorrunning{K. Kotera, F. Mottez, G. Voisin, J. Heyvaerts}
\date{\today}
\author{Kumiko Kotera \inst{1,2}, Fabrice Mottez \inst{3}, Guillaume Voisin \inst{3}, Jean Heyvaerts \inst{4}\thanks{Deceased July 2013.}}

   \institute{Sorbonne Universit\'es, UPMC Univ. Paris 6 et CNRS, UMR 7095,  Institut d'Astrophysique de Paris, 98 bis bd Arago, 75014 Paris, France 
   \and
   Department of Astronomy \& Astrophysics, Kavli Institute for Cosmological Physics, The
  University of Chicago, Chicago, IL 60637, USA.
              \and
             LUTH, Observatoire de Paris, PSL Research University, CNRS, Unviersit\'e Paris Diderot,
              5 place Jules Janssen, 92190 Meudon, France.
                           \and
             Observatoire Astronomique, Universit\'e de Strasbourg,
                     11, rue de l'Universit\'e, 67000 Strasbourg, France.} 
                     
\abstract{}{We investigate the evaporation of close-by pulsar companions, such as planets, asteroids, and white dwarfs, by induction heating.} {Assuming that the outflow energy is dominated by a Poynting flux (or pulsar wave) at the location of the companions, we calculate their evaporation timescales, by applying the Mie theory.}{ Depending on the size of the companion compared to the incident electromagnetic wavelength, the heating regime varies and can lead to a total evaporation of the companion. In particular, we find that inductive heating is mostly inefficient for small pulsar companions, although it is generally considered  the dominant process.}{ Small objects like asteroids can survive induction heating for $10^4\,$years at distances as small as $1\,R_\odot$ from the neutron star. For degenerate companions, induction heating cannot lead to evaporation and another source of heating (likely by kinetic energy of the pulsar wind) has to be considered. It was recently proposed that bodies orbiting pulsars are the cause of fast radio bursts; the present results explain how those bodies can survive in the pulsar's highly energetic environment.}

\begin{document}

\maketitle

\section{Introduction}

Pulsars are highly magnetized, rapidly rotating neutron stars (NS) that lose their energy principally via electromagnetic cooling, which results in their spin-down. The pulsar outflow comprises a low-frequency ($\omega=2 \pi/ P$, with $P$ the pulsar spin period) Poynting flux-dominated component (also called the pulsar wave), a relativistic wind, and high-energy radiation from the magnetosphere.
The energy of the outflow is believed to be dominated by Poynting flux close to the star, and by relativistic particles farther out (as observed in particular in the case of the Crab pulsar). However, the location of this transition remains a puzzle to the community (see, e.g., \citealp{Kirk09} for a review).

Pulsars are often observed to evolve in multiple systems, and evaluating the amount of energy absorbed by the companions is paramount to understanding their evolution and emission.
The most common companions are white dwarfs \citep{Savonije87,Manchester05,Ransom14} or M-dwarfs \citep{Roberts13}, but smaller objects such as planets, asteroids, or comets can also orbit pulsars \citep{Wolszczan92,Thorsett93,Bailes11}. The existence of asteroid belts around pulsars has been invoked by several authors to explain timing irregularities \citep{Shannon13}, anti-glitches \citep{Huang14}, or burst intermittency \citep{Cordes08,Deneva09,Mottez13}. 

Asteroids around pulsars could also be at the origin of fast radio bursts (FRBs). These brief radio signals (typically 5 ms at a given frequency) are dispersed in frequency (as for pulsar signals), but with a dispersion corresponding to cosmological distances (e.g., \citealp{Champion16} and references therein). Going against the mainstream scenarios involving the collapse of massive objects, \cite{Mottez14} proposed that FRBs could be emitted by bodies orbiting extragalactic pulsars (see also \citealp{Dai16}). Highly collimated waves would be produced by the magnetic wake of these objects immersed in the relativistic pulsar wind. The collimation would enable their detection even at distances of  hundreds of Mpc. This model can naturally produce the repeating bursts reported by \cite{Spitler16} in the presence of an asteroid belt.

For massive companions, the standard scenario stipulates that old neutron stars are spun up to periods $<10\,$ms in close binary systems by transfer of mass and angular momentum by the low-mass companion \citep{Alpar82}. This accretion process leads to the production of X-rays detected as low-mass X-ray binaries. 
In particular, the so-called black widow (dwarf companion with mass $m\sim 0.002-0.07\,M_\odot$) and redback (M-dwarf companion with mass $m\sim 0.1-0.4\,M_\odot$) pulsars illustrate the importance of energy absorption by companions on the evolution of binary systems. These systems have very low-mass secondaries and orbital periods of less than 10\,h. Early studies pointed out that these pulsars should theoretically evaporate their companion by high-energy radiation \citep{Ruderman89} and/or by Roche lobe overflow. At the time of their discovery, it was assumed that black widows provided the missing link between low-mass X-ray binaries and isolated millisecond pulsars.

Since pulsar environments can evaporate white dwarves and M-stars orbiting them, it may seem surprising that smaller planets and asteroids could be maintained in a solid state at similar distances to the neutron star. In the present paper, we propose {an explanation for their survival.}   

{ To date, it has been confirmed that two pulsars (PSR B1257+12 and PSR B1620-26) host planets. PSR B1257+12 hosts three planets at distances on the order of the astronomical unit (in the pulsar wind) with orbital periods in the range of days and weeks \citep{Wolszczan92}. The distances are therefore greater than in close binary systems (orbital periods in the range of hours) such as black widow and redback systems, which can provide an explanation to the stability of the planets. PSR B1620-26 is a neutron star-white dwarf binary, that was confirmed to host a Jovian mass companion orbiting at 23 AU \citep{Sigurdsson03}. The distant orbit makes this system ill-suited for our framework, and could explain by itself why the companion has not been evaporated. On the other hand and if confirmed,  the suspected belt of asteroids at a small distance from PSR 1931+24, with an orbital period ranging in minutes \citep{Cordes08,Deneva09,Mottez13_a,Mottez13} is puzzling and would require an alternative justification for its survival. PSR J1719-1438 presents an orbiting candidate planet \citep{Bailes11}. Its Jovian mass companion has a short orbital period of $\sim 2.2\,$h, implying a compact orbital distance of $\sim 0.95\,R_{\odot}$, and a minimum mean density of $23.3\,$g\,cm$^{-3}$. This density suggests that it may be an ultra-low-mass carbon white dwarf rather than a planet. The evaporation timescale of this companion should also be investigated. 
} 
 
At distances within a few light-cylinder radii from the neutron star, there is a general agreement that most of the energy outflowing from the pulsar is under the form of the Poynting flux associated with the pulsar wave \citep{Kirk09}. The pulsar wave is the electromagnetic structure created by the rotation  of the neutron star magnetic field at the pulsar spin frequency (with an inclination angle $i \ne 0$ relative to the spin axis).  Outside the light-cylinder, this corresponds to a low-frequency spherical wave propagating at a velocity $ \sim c$ and of wavelength equal to two light-cylinder radii \citep{Deutsch_1955}. On the scale of a companion radius, this wave can be considered  a plane wave.  
Because the pulsar wave constitutes the main flux of energy out of the pulsar, we consider that it is potentially the main source of heat for pulsar companions. In the remaining sections, we investigate the physical parameters that control the efficiency of the pulsar companion heating by the pulsar wave.

\section{Companion heating by induction: model}

All numerical quantities are denoted $Q_x\equiv Q/10^x$ in cgs units unless specified otherwise.

\subsection{Energy flux of the pulsar wind}

The energy loss rate (or luminosity) of a pulsar with moment of inertia $I$, rotation period $P$, radius $R_\star$, dipole magnetic field strength $B_\star$, and corresponding period derivative $\dot{P}$ reads (e.g., \citealp{Shapiro83})
\begin{eqnarray}\label{eq:Edot}
L_{\rm p} \equiv \dot{E}_{\rm rot} &=& \frac{8\pi^4 R_\star^6 B_\star^2}{3c^3 P^4} = I(2\pi)^2\frac{\dot{P}}{P^3}\\
&\sim& 3.9\times 10^{35}\,{\rm erg\, s}^{-1}\,I_{45}\dot{P}_{-20}P_{-3}^{-3}\ \nonumber\\
&\sim& 9.6\times 10^{34}\,{\rm erg\, s}^{-1}\,I_{45}P_{-3}^{-4}B_{\star,8}^2R_{\star,6}^6\ .\nonumber
\end{eqnarray}
The energy flux in the pulsar wind at distance $a$, large compared to the pulsar light cylinder radius, $R_{\rm L} = cP/(2\pi)\sim 4.8\times 10^8\,{\rm cm}\,P_{-3}$, can then be written \citep{Arons93}
\begin{eqnarray}\label{eq:FwEr}
F_{\rm w} &=&\frac{L_{\rm p}}{4\pi f_{\rm p}a^2}
= {  \frac{1}{f_{\rm p}a^2}\frac{ 2 \pi^3 R_\star^6 B_\star^2}{3c^3 P^4} }
=\frac{I\pi}{f_{\rm p}a^2}\frac{\dot{P}}{P^3}\\
&\sim& 1.8\times 10^{13}\,{\rm erg\,s}^{-1}{\rm cm}^{-2}\,f_{\rm p}^{-1}\,\left(\frac{a}{R_\odot}\right)^{-2}I_{45}P_{-3}\dot{P}_{-20} \ ,\nonumber
\end{eqnarray}
where  $f_{\rm p}=\Delta\Omega_{\rm p}/(4\pi)$ is the fraction of the sky into which the pulsar wind is emitted. We note that this flux can also be expressed as
$F_{\rm w} ={cB^2(1+\sigma_B)}/({4\pi\sigma_B})$;
 the magnetization parameter is defined as $\sigma_B\equiv B^2/[4\pi c^2(n_{\rm i}m_{\rm i}+\kappa_\pm m_\pm n_\pm)]$, i.e., the ratio of the magnetic energy flux to the kinetic energy flux in the comoving wind frame. Here $B$, $n_{\rm i}$, $m_{\rm i}$, and $n_{\pm}$ and $m_{\pm}$ are respectively the magnetic field strength, the number density, the mass of ions, and pairs of the cold plasma, and $\kappa_\pm$ the pair multiplicity. In this work, we assume that the wind is Poynting-flux dominated ($\sigma_B\gg 1$) in the region where the companion is located.  It can be demonstrated that this is valid out to the termination shock for a supersonic, radially expanding ideal magneto-hydrodynamics (MHD) wind. In more realistic situations, the dissipation to kinetic energy could happen earlier, but all models predict a Poynting-flux dominated wind close to the star (e.g., \citealp{Kirk09}).

The companion at a distance $a$ can intercept a fraction $f$ 
of this flux, provided that it falls in the wind beam and absorbs a flux, $F_{\rm abs} = f Q_{\rm abs} F_{\rm w}$,
{which can be written in terms of absorbed luminosity}:
\begin{equation}\label{eq:Labs}
L_{\rm abs} = \pi R^2 f Q_{\rm abs} F_{\rm w} = \frac{2 \pi^4}{3c^3}Q_{\rm abs} \frac{ f }{f_p}\frac{R_\star^6 B_\star^2}{P^4}.
\end{equation}
Here, $Q_{\rm abs}$ is the energy absorption efficiency. Using the magnetization parameter $\sigma$ defined above, this quantity can be decomposed into two components: the absorption of energy from the pulsar wave, characterized by an efficiency $Q_{\rm em}$, and the absorption of kinetic energy from the pulsar wind, of efficiency $Q_{\rm kin}$,
\begin{equation}\label{eq:Qabs_decomp}
Q_{\rm abs}= \frac{\sigma_B}{1+\sigma_B}\, Q_{\rm em} + \frac{1}{1+\sigma_B} \, Q_{\rm kin} \ .
\end{equation}

The next two sections are devoted to the evaluation of $Q_{\rm em}\sim Q_{\rm abs}$ for a Poynting-flux dominated case when $\sigma_B\gg 1$.

\subsection{Absorption properties of a body in an electromagnetic wave} \label{sec_Q_abs}
The absorption properties of a spherical body in an electromagnetic wave can be modeled by the Mie theory \citep{Hulst_81}. Two parameters govern the regime in which absorption or scattering occurs: the ratio of the size of the body to the incident wavelength 
\begin{equation} \label{def_x}
x\equiv\frac{R}{cP} = \frac{R}{2 \pi R_L} \sim 3.3\times 10^{-3}\,R_5P_{-3}
\end{equation} 
and the complex refractive index of the medium $N=N_{\rm r} + N_{\rm i}$, where $N_{\rm r}$ and $N_{\rm i}$ are real.

\subsubsection{Refractive index of the companion} \label{section:refrac}
 
 Companions orbiting a 10 ms pulsar at a distance $a=1\,R_{\odot}$ see an alternating magnetic field of amplitude  $B \le 10$ G with a  100 Hz frequency\footnote{The wave amplitude is much smaller than in a domestic transformer (where magnetic fields of $10^4$ G are common).}. 
 The electric field induced by the pulsar wind ($v \sim c$)  is $E \le 10^5$ V/m. This is less than the electric strength encountered in most materials (larger than $10^6$ V/m). Therefore, the companions are not ionized  by the electric field induced by the pulsar wave.
 Similar conditions are met for companions orbiting standard pulsars ($P \sim 1$ s) beyond the light cylinder. Therefore, the usual approximations and textbook data can be used to estimate the electromagnetic properties of the constituents of pulsar companions.

 The refractive index can be calculated using the Maxwell equations and the material equation $\vec J= \sigma \vec E$ where $J$ is the current density and $\sigma$ is a scalar conductivity \citep{Born_1980}. The dispersion equation reads  
 \begin{equation}
 \mu \epsilon \omega^2 - i 4 \pi \omega \mu \sigma + k^2 c^2 =0,
  \end{equation} 
  where $\omega=2\pi/P$, $k$ is the wave number, and $\epsilon$ and $\mu$ the dielectric permittivity and magnetic permeability of the medium, respectively. The refractive index $N$ defined by
  $N^2\equiv k^2 c^2 / \omega^2$ can be written 
   \begin{equation}\label{eq:N2}
  N^2 = \mu \left(  \epsilon  - i \frac{4\pi  \sigma}{\omega} \right).
  \end{equation} 
  
Pulsars rarely spin faster than with a millisecond period, thus $\omega \lesssim 6 \times 10^3$ s$^{-1}$ (a strict lower bound on the spin is given by \citealp{Haensel99}). For low electromagnetic frequencies (up to infrared frequencies), $\sigma$ is real in any kind of material. For companions made of degenerate matter, the electrical conductivity can be estimated as $\sigma\sim 10^{24}\,{\rm s}^{-1}\, \rho^{2/3}T^{-1}$, for densities ranging from $\rho\sim 10^{6-12}\,$g\,cm$^{-3}$, and $T$ the stellar temperature \citep{Canuto70}. White dwarfs typically have $\rho\sim 10^9\,$g\,cm$^{-3}$ and $T\sim 10^{4-7}\,$K. For $T=10^5$ K, $\sigma  \sim 10^{22}$ s$^{-1}$. To date, the planets orbiting pulsars that have been discovered could be either low-mass white dwarfs made of degenerate matter, or made of ordinary matter. For $T=10^5$ K, $\sigma \sim 10^{25}$\,s$^{-1}$. Smaller objects such as asteroids could be modeled as silicate rock or as ferrous metals. The conductivity of sea water is on the order of $\sigma \sim 10$\,Mho m$^{-1} \sim 10^{10}$ s$^{-1}$. In metals, $\sigma \sim 10^6 - 10^7$\,Mho m$^{-1} \sim 10^{17}$\,s$^{-1}$. For solid rocks in the terrestrial crust, $\sigma \sim 10^{-3}$\,Mho m$^{-1} \sim 10^{6}$\,s$^{-1}$, but in some conductive belts (for instance between the African Rift and Namibia)  $\sigma \sim 10^{-1}$\,Mho m$^{-1}$ \citep{Gough_1989}. 

In all materials, the permittivity $\epsilon$ is of order unity, rarely exceeding 100. For instance, water-free iron ore has $\epsilon \sim 5$ and iron oxides $\epsilon \sim 2$. 
        
Given the above estimates, one can neglect the real part of $N^2$ in our framework. The real and imaginary parts of $N$ can then be extracted from Eq.~(\ref{eq:N2}):
\begin{equation} \label{eq:N2bis}
N_{\rm r}=-N_{\rm i} = \sqrt{\frac{2 \pi \sigma \mu}{\omega}}=\sqrt{\sigma \mu P}. 
\end{equation}
For example, for a companion made of iron (or any other ferromagnetic material), $\mu > 10^3$ and $N_{\rm r} \gg 10^3$. For solid rocks in the terrestrial crust, $\mu \sim 1$ and $N_{\rm r} \sim {10^3}$ for a $P=1$\,s pulsar and $N_{\rm r} \sim 30$ for a $P=1$\,ms pulsar.  

For white dwarfs, the very large conductivity $\sigma$ implies $N_{\rm r} \gg 10^6$.

\subsubsection{Absorption coefficient}

Here we compute  the absorption coefficient $Q_{\rm abs}$, defined in Eq.~(\ref{eq:Labs}) as the proportion of energy absorbed by the companion, relative to the flux of incident energy of the wave through the section of the companion. 
We perform a numerical computation of $Q_{\rm abs}$ using a version of the {\tt Damie} code, based on \cite{Lentz_1976}\footnote{
http://diogenes.iwt.uni-bremen.de/vt/laser/codes/ddave.zip}.
We have tested the program with real and complex values of $N^2$, and compared the results with analytical approximations given in \cite{Hulst_81} for various regions of the $(x, N)$ parameter space. 
 We find that numerical and analytical methods give consistent values within an error of $<21\%$. This level of  agreement is sufficient for our purpose as we are mostly interested in order-of-magnitude estimates, given the uncertainties of our other parameters.

\begin{figure}
\vspace*{2mm}
\begin{center}
\includegraphics[width=\columnwidth]{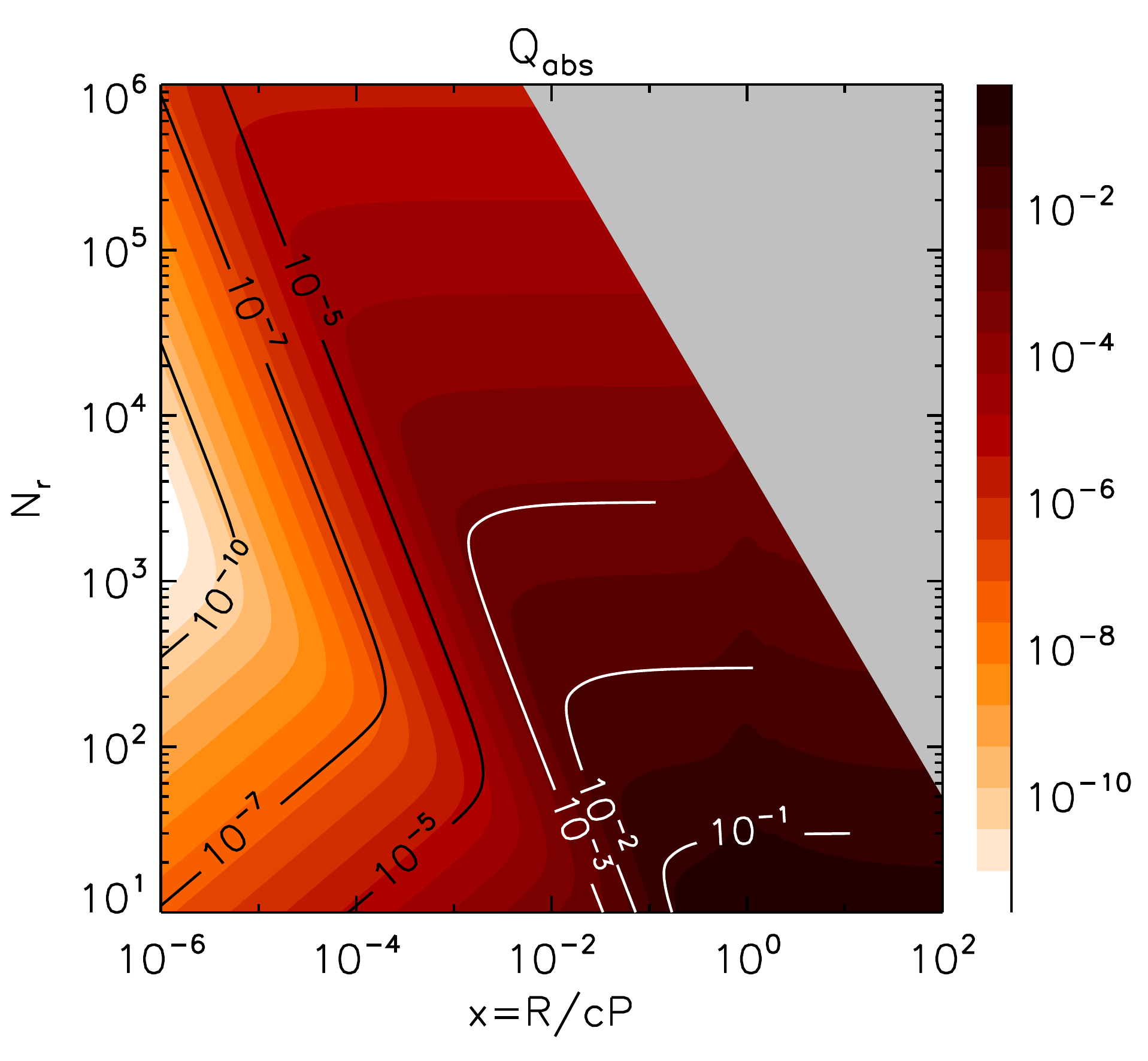}
\end{center}
\caption{Absorption coefficient $Q_{\rm abs}$ as a function of the size ratio $x=R / c P$ and of the refractive index $N_{\rm r}$ of the companion in the Metallic regime ($N=N_{\rm r}=-N_{\rm i}$). Overlayed are the contours of the analytical approximation, in black for Region 1 ($xN_{\rm r}\ll 1$, Eq.~\ref{eq:Qabs1}) and in white for Region 2 ($xN_{\rm r}\gg 1$, Eq.~\ref{eq:Qabs2}).
The gray-shaded region corresponds to parameters outside the confidence range of the computation with {\tt Damie}. }
\label{fig_Q_abs_metalic}
\end{figure}

 Figure \ref{fig_Q_abs_metalic} shows the value of the absorption coefficient computed with {\tt Damie} as a function of the size parameter $x=R/(cP)$ and of the refractive index $N$, in the so-called Metallic regime, where $N=N_{\rm r}=-N_{\rm i}$ for values of $N$ larger than one. 

As expected, for large objects ($x \gtrsim 1$), $Q_{\rm abs} \sim 1$. This means that all the flux received from the pulsar wave is absorbed. The values  $x \gtrsim 1$ correspond to planet-sized and white dwarf-sized objects ($R \ge 1000$ km) in a millisecond pulsar wind ($c P =3000$ km for $P=10$ ms). For such objects, taking into account the Mie theory does not modify the heating rate. 

On the other hand, for smaller objects with $x<1$, the level of absorption is reduced by orders of magnitude. 
For instance, for a kilometer-sized asteroid orbiting a standard pulsar ($P=1$\,s), $x=3\times10^{-6}$. If it is made of rocks, $N_{\rm r} \sim 30$, and it can be read $Q_{\rm abs} \sim 10^{-8}$. A value $Q_{\rm abs} \sim 10^{-12}$ is even reached when $N_{\rm r} \sim 10^3$. 

For $x\ll 1$, \cite{Hulst_81} gives an analytical expression for the absorption coefficient (we note, however, the corrected typo in the numerator)
\begin{equation}\label{eq:Qabs2}
Q_{\rm abs} = \frac{3}{N_{\rm r}}\left[ \frac{\sinh (2xN_{\rm r}) + \sin (2xN_{\rm r})}{\cosh (2xN_{\rm r}) - \cos (2xN_{\rm r})} -\frac{1}{xN_{\rm r}}\right]\ .
\end{equation}
The absorption coefficient computed with Eq. (\ref{eq:Qabs2}) is plotted in Fig.~\ref{fig_Q_abs_metalic} (white contours). 
For small values of $x$, this approximation is numerically unstable. 
An approximation for the parameter space where $N_{\rm r} x\ll 1$ called ``Region 1'' was proposed \citep{Hulst_81}:
\begin{equation}\label{eq:Qabs1}
Q_{\rm abs} = \frac{12 x }{N_r^2} +\frac{2 x^3 N_r^2}{15} \ .
\end{equation}
As shown in Fig.~\ref{fig_Q_abs_metalic} (black contours), this approximation is valid for the smallest values of $N_{\rm r} x$.
The domain corresponding to $xN_{\rm r}\ge 1$ was denoted ``Region 2'' by \cite{Hulst_81}.
For $xN_{\rm r}\gg 1$, Eq.~(\ref{eq:Qabs2}) can be approximated as
\begin{equation}\label{eq:Qabs_approx2}
Q_{\rm abs} = \frac{3}{N_{\rm r}}\ .
\end{equation}Region 2 corresponds roughly to planets and white dwarfs, while Region 1 corresponds to planetesimals, asteroids, and comets\footnote{From an electromagnetism point of view, one can understand these two regions as different regimes of penetration of the magnetic field inside the object. In Region 2, the inductive currents do not penetrate deep under the surface (the resistive skin-depth is shallow). For Region 1, the skin-depth is approximately the entire size of the object, and the magnetic field can be considered as roughly uniform over it. }. 

White dwarfs, with $N_{\rm r}\gg 10^6$, are hence highly reflective bodies, as confirmed by the Mie theory. Indeed, $Q_{\rm abs} \rightarrow 0$ (Eq.~\ref{eq:Qabs_approx2}). If the pulsar outflow is under the form of a pulsar wave while it reaches the white dwarf, it should not absorb any energy and should not evaporate. This goes against the existence of black widow systems. It thus seems likely that the outflow is not Poynting-flux dominated as it reaches the white dwarf companion. One possible scenario is that the pulsar wave energy is dissipated by plasma heating, in the dense plasma surrounding the evaporating white dwarf.

\section{Effects on the evaporation of companions} \label{sec_t_ev}

We discuss in this section the impact of $Q_{\rm abs}$ on the evaporation rate of pulsar companions.
A simple energy balance gives the evaporation timescale of the companions as
\begin{equation}
        {E_\mathrm{abs}(t_{\rm ev})}={[E_{\rm g} + E_{\rm c} - E_\mathrm{rad}](t_{\rm ev})} \ ,
\end{equation}
where $t_{\rm ev}$ is the minimum time of evaporation, $E_\mathrm{abs}$ is the total energy absorbed  at time $t$, $E_\mathrm{rad}$ the total radiated energy, $E_{\rm g}$ the gravitational binding energy, and $E_{\rm c}$ the cohesive energy\footnote{Cohesive energy is  the energy required to form separated neutral atoms in their ground electronic state from the solid at a given temperature under a given pressure \citep{Kittel_issp}.}. 
        Two regimes will be distinguished in this paper. For massive, degenerate companions the cohesive energy can be neglected (section \ref{sec_massivecomp}), while for low-mass companions it dominates (section \ref{sec_lightcomp}). 
        Gravitational and cohesive energies are of the same order of magnitude for bodies about the mass of the Earth. 
        In both cases, we neglect the radiative cooling, but we will include it in a more refined treatment in a future paper.

\subsection{Ablation and evaporation of massive companions (degenerate and non-degenerate)}\label{sec_massivecomp}

\begin{figure*}[th]
\begin{center}
\includegraphics[height=0.3\textwidth]{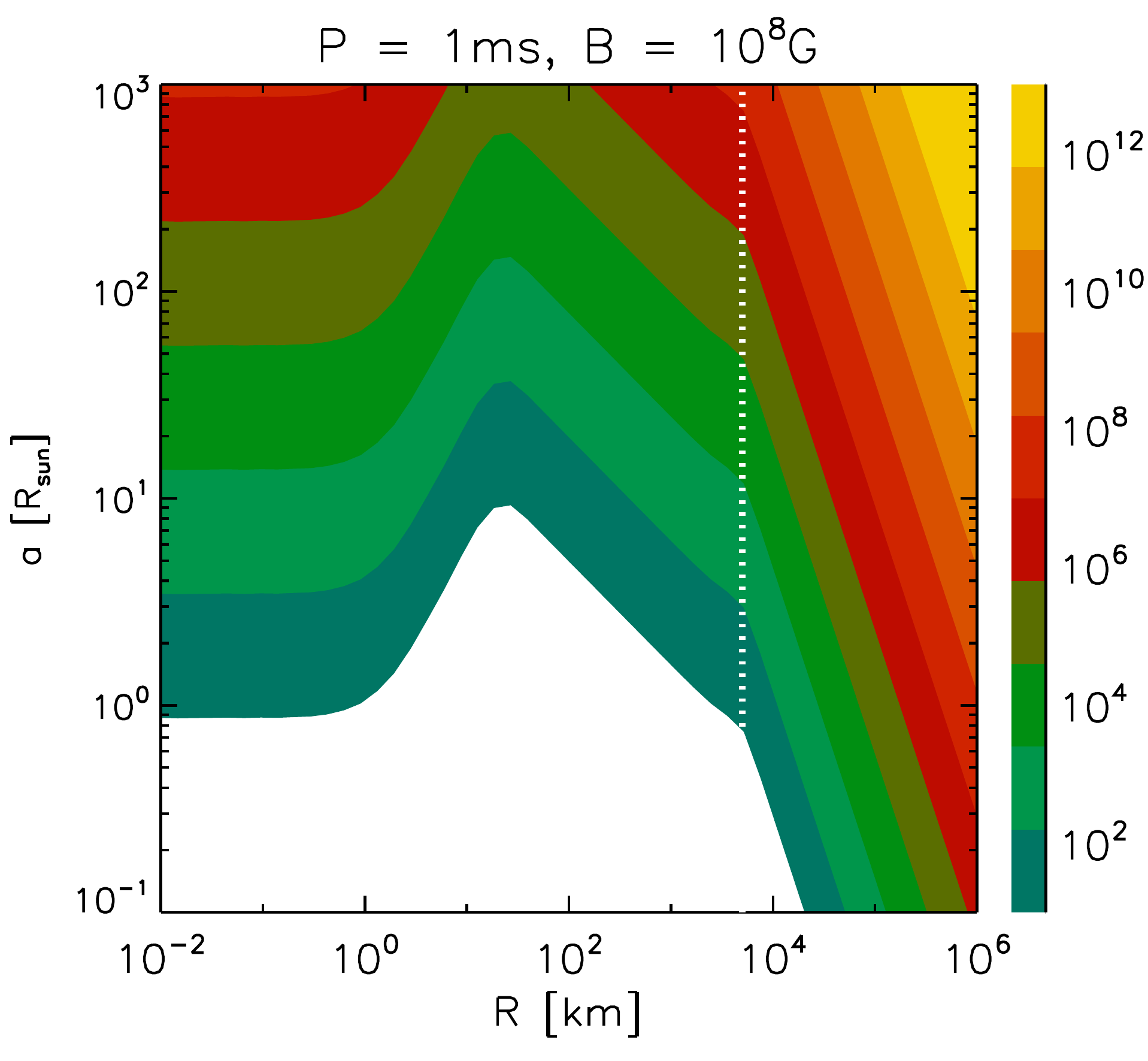}
\includegraphics[height=0.3\textwidth]{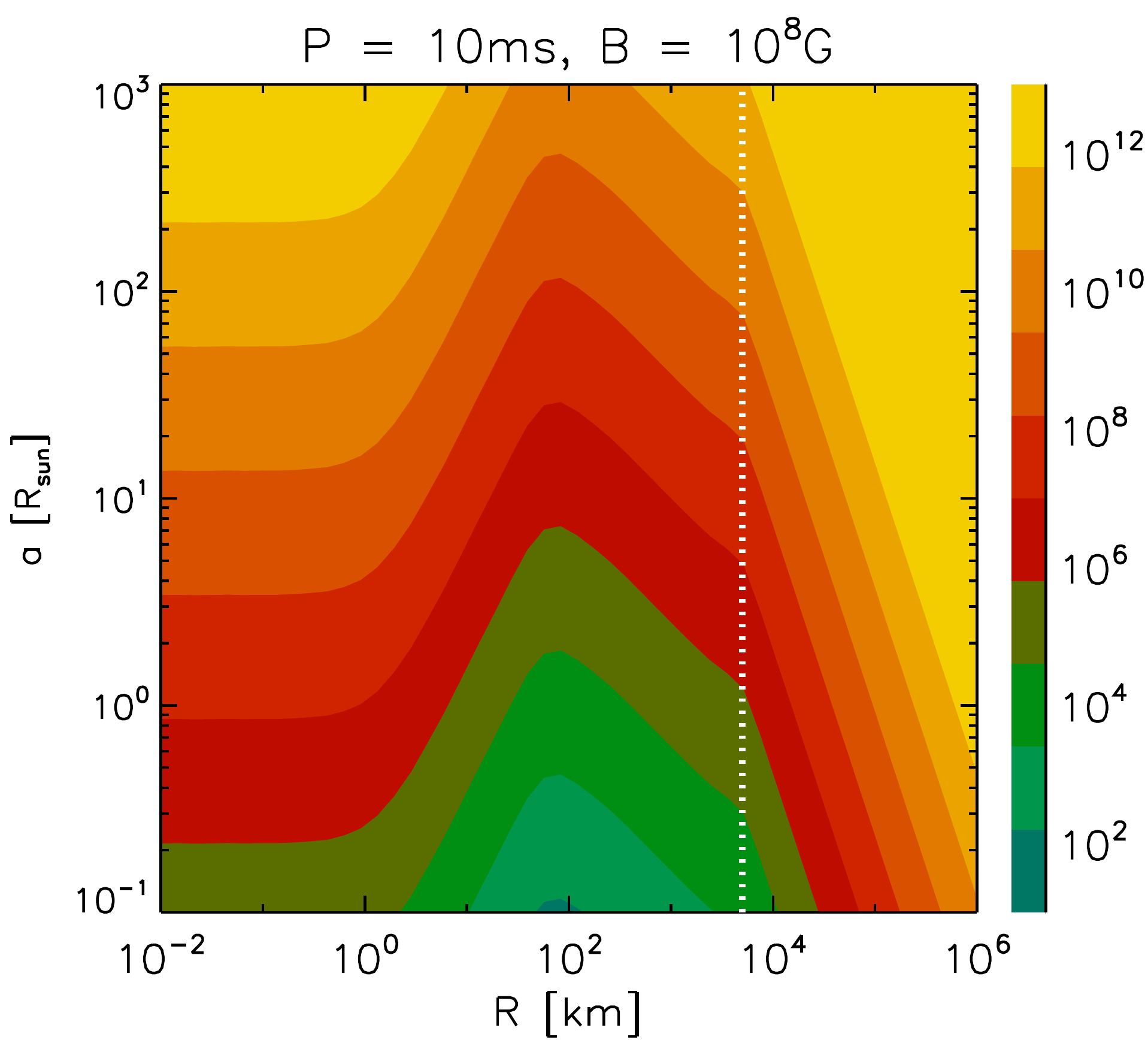}
\includegraphics[height=0.3\textwidth]{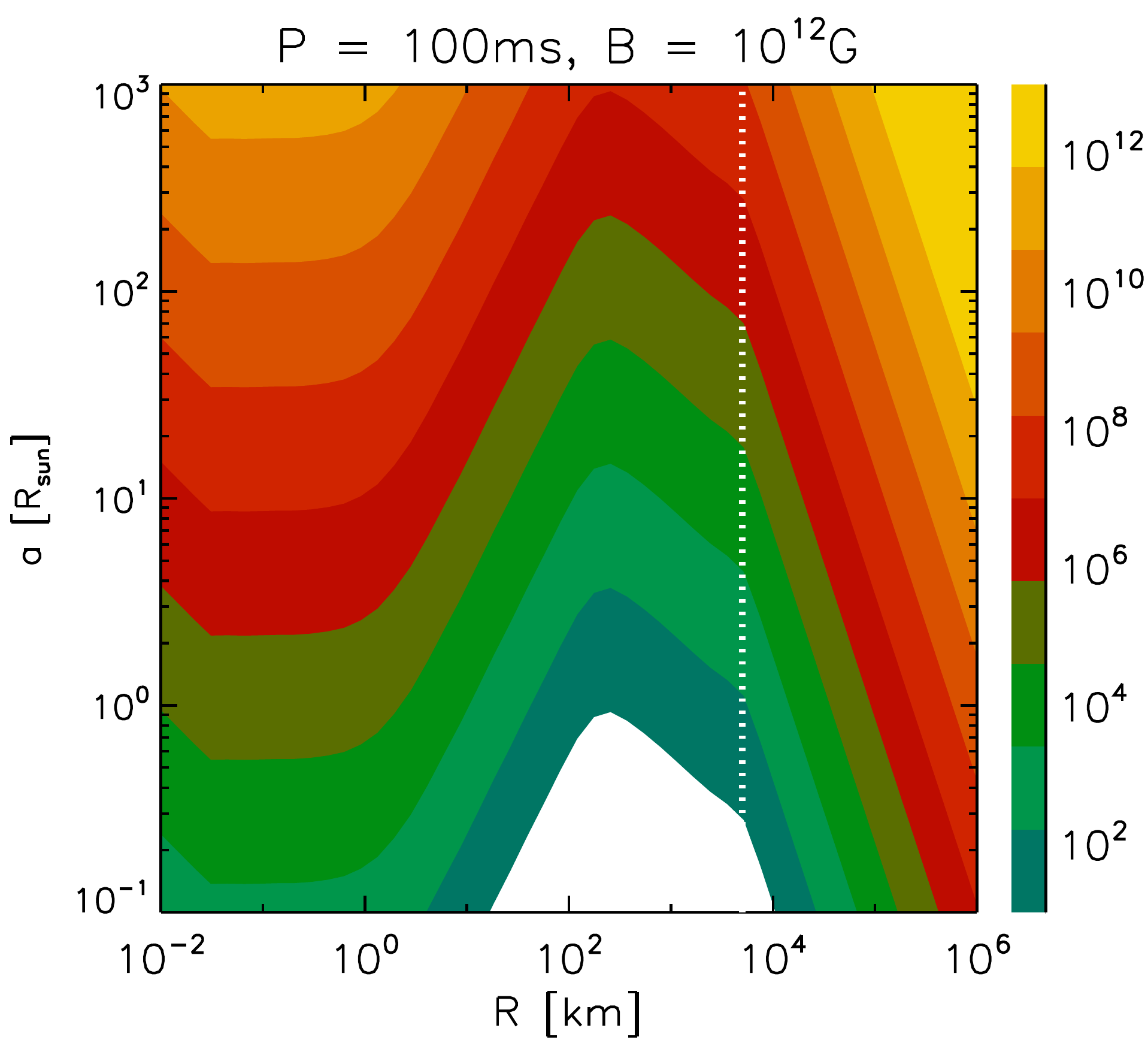}
\end{center}
\caption{Contours of evaporation timescales (in yrs) of non-degenerate pulsar companions ($t_{\rm ev,g}$ and $t_{\rm ev,c}$, Eqs.~\ref{eq:tevg} and \ref{eq:tevc}) as a function of the companion radius $R$ and its distance $a$ to the pulsar in solar radii, for pulsar spin periods and magnetic fields as indicated (left to right: $P=1, 10, 100\,$ms and $B=10^8,10^8,10^{12}\,$G), for conductivity times magnetic permeability $\sigma\mu=10^6\,$s$^{-1}$ and density $\rho = 1\,$g\,cm$^{-3}$. The white dotted line delimits the region where the cohesive energy dominates over gravitational energy (for small companions) for evaporation. The white region indicates evaporation times $t_{\rm ev}< 10\,$yrs and yellow regions $t_{\rm ev}> 10^{12}\,$yrs. }
\label{fig:tev}
\end{figure*}

For massive companions, the dominant process for evaporation is the gravitational escape of the gas. 
The evaporation timescale, $t_{\rm ev}$, results from the balance between the energy flux input from the pulsar wave and the rate of gravitational escape of the heated material \citep{vandenHeuvel88}. The luminosity available to drive the wind of a companion of mass $M$ and radius $R$ located at a distance $a$ from the pulsar is then $(1/2)\dot{M}v^2 = Q_{\rm abs}(f/f_{\rm p})(R/2a)^2L_{\rm p}$. Assuming that the wind velocity is on the order of the escape velocity $v = (2GM/R)^{1/2}$ from the surface of the companion, the gravitational binding energy of the companion  reads 
\begin{equation}
E_{\rm g} = \frac{GM^2}{R} \sim 3.8\times10^{48}\, {\rm erg} \,\left(\frac{M}{M_\odot}\right)^2\frac{ R_\odot}{R} \ ,
\end{equation}
and the evaporation timescale
\begin{eqnarray}
t_{\rm ev} \equiv \frac{M}{\dot{M}}&=&\frac{4GM^2a^2}{R^3}\frac{f_{\rm p}}{f}(L_{\rm p}Q_{\rm abs})^{-1} \\
&=& \frac{3c^3}{2\pi^4}\frac{P^4}{R_\star^6 B_\star^2}\frac{GM^2}{R^3}\frac{a^2 f_{\rm p}}{f Q_{\rm abs}}\ .
\end{eqnarray}
For low values of $M$ when the star is degenerate, the mass-radius relation follows roughly $R/R_\odot = 0.013(1+X)^{5/3}(M/M_\odot)^{-1/3}$, where $X$ is the hydrogen fraction;  the evaporation time can then be expressed
{ \begin{eqnarray} \label{tev_ordinaire_white_dwarf}
\frac{t_{\rm ev,WD}}{{\rm yr} }= 2.3 \times 10^{12}\, ( {1+X})^{-5}\left( \frac{M}{M_\odot}\right)^{3} 
\left( \frac{a}{R_\odot}\right)^{2} \times\nonumber \\
\frac{f_{\rm p}}{f}\frac{1}{L_{\rm p,35}\,{Q_{\rm abs}}} \ .
\end{eqnarray}
The spin-down luminosity is on the order of $L_{\rm p}\sim10^{35}\,$erg\,s$^{-1}$ for a typical millisecond pulsar with $R_\star=10$\,km, $P=10$\,ms, $B=10^8$\,G, and on the order of $L_{\rm p}\sim10^{31}\,$erg\,s$^{-1}$ for a standard pulsar with $P=1000$\,ms, $B=10^{12}$\,G (see Eq.~\ref{eq:Edot}).

This equation is analogous to those in  \cite{vandenHeuvel88}.
These estimates can be applied to PSR B1957+20, a benchmark black widow system suspected of   ablating its companion (e.g., \citealp{Phinney88,Kluzniak88,Arons93,Callanan95,Khechinashvili00,Stappers03,Reynolds07,vanKerkwijk11}). PSR B1957+20 has a period of $P=1.61\,$ms, and its dwarf companion of mass $M=0.021\,M_\odot$ 
orbits at a distance $a=2.1\,R_\odot$ ($P_{\rm orb}=9.2$ hr), hence $t_{\rm ev,WD}\sim {\cal O} (10^6\,Q_{\rm abs}^{-1})\,$yrs. However, as argued at the end of Section~\ref{section:refrac}, the high refractive index of white dwarfs implies that $Q_{\rm abs}\ll 1$, thus $t_{\rm ev,WD}\gg10^6\,$yrs. The conclusions derived in \cite{vandenHeuvel88}, namely that the evaporation of the companion (strongly dependent on $R_\star$ and $X$) takes only a few million years, remain valid only if the absorbed energy is principally kinetic (Eq.~\ref{eq:Qabs_decomp}). The same conclusions apply for PSR J1719-1438 (period $P=5.7\,$ms), if its companion is indeed an ultra-low-mass carbon white dwarf with $M= 0.015\times 10^{-3}\,M_\odot$, $R=4.2\times 10^4\,$km, and $a=0.95R_\odot$.

For planet companions made of ordinary matter, the mass/radius relationship simply reads $M =(4 \pi /3) R^3 \rho$, where $\rho$ is the average density. The evaporation time can then be expressed
\begin{eqnarray} \label{eq:tevg}
\frac{t_{\rm ev,g}}{\rm yr}= 7.2 \times 10^{-12}  \left(\frac{R}{\rm km}\right)^3 \left( \frac{\rho}{{\rm \,g\,cm}^{-3}}\right)^{2} \left( \frac{a}{R_\odot}\right)^{2}\times\nonumber \\
 \frac{f_{\rm p}}{f}\frac{1}{L_{\rm p,35}\,{Q_{\rm abs}}}\ .
\end{eqnarray}
This equation can be applied to the planets orbiting PSR 1257+12. 
The results are given in Table \ref{table_temps_evaporation_3} for two values of the refractive index $N_{\rm r}$ that correspond to planets made of rocks and to more metallic planets. We find that the closest planet can survive a few million years. The two other planets can reach a billion years. This estimate neglects the thermal energy radiated by the planet; therefore,  the evaporation time through Poynting flux absorption may be underestimated. 

If the companion of PSR J1719-1438 is actually a non-degenerate planet with minimum mean density $\rho=23.3\,$g\,cm$^{-3}$, then $x\gg 1$, but with $N_{\rm r}\lesssim 100$, implying $Q_{\rm abs}\gtrsim 10^{-2}$ and $t_{\rm ev,g}={\cal O}(10^{6-8}\,{\rm yrs})$ for $f=f_{\rm p}=1$, which is compatible with the observation of the companion.

\begin{figure}[h]
\begin{center}
\includegraphics[height=0.45\textwidth]{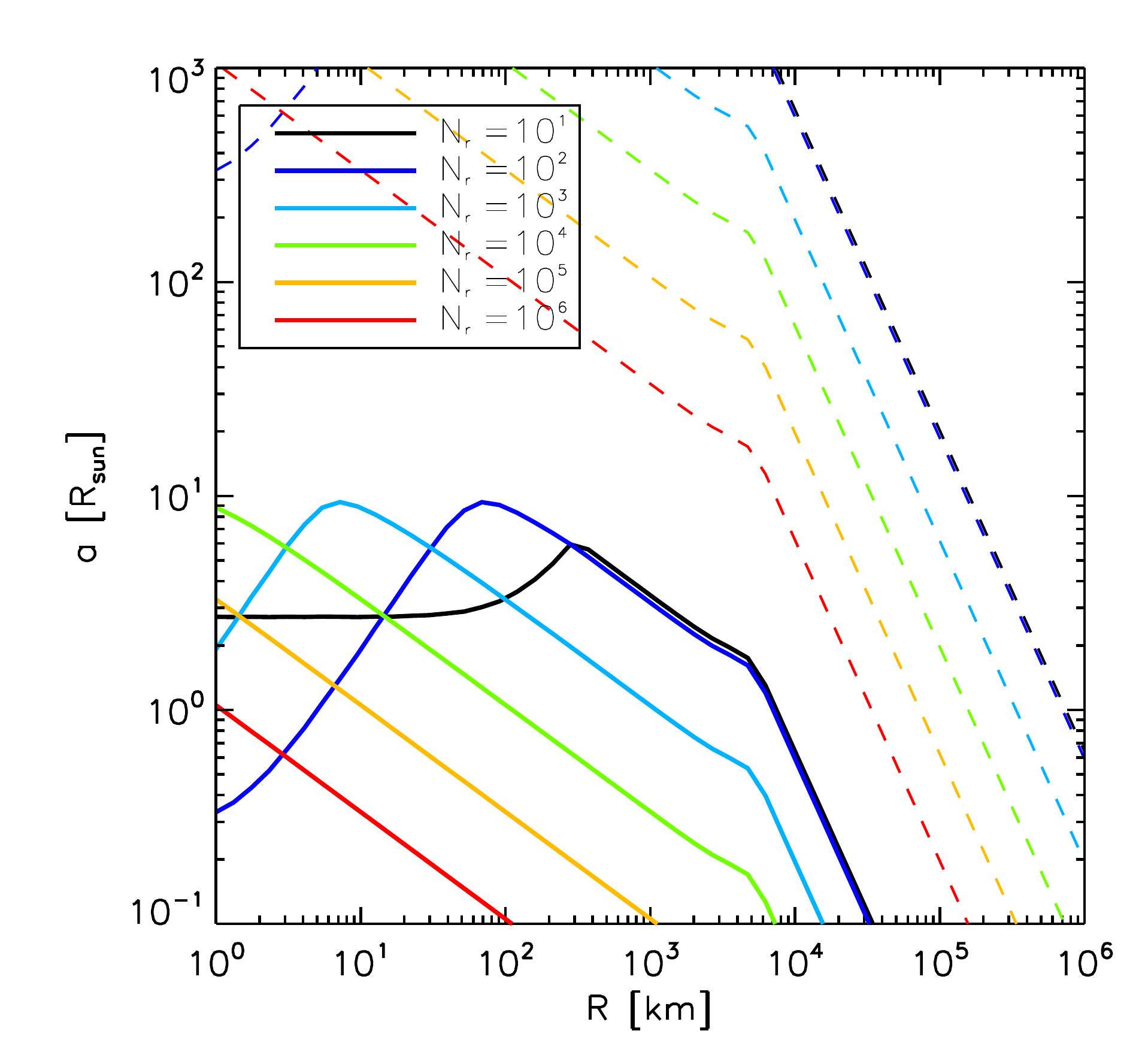}
\end{center}
\caption{Influence of companion composition on evaporation timescale. The contours of $t_{\rm ev}=10^6\,$yrs (solid lines) and $t_{\rm ev}=10^{12}\,$yrs (dashed lines) are represented as a function of companion distance $a$ and size $R$ for $P=10\,$ms, $B=10^8\,$G, and an indicative (and conservative) density $\rho = 1\,$g\,cm$^{-3}$, and  for various companion refractive indices as indicated in the legend (increasing from right to left): $\sigma\mu=10^{4-14}\,$s$^{-1}$, corresponding to $N_{\rm r}=10^{1-6}$.}
\label{fig:tevNr}
\end{figure}

\begin{landscape}
\begin{table}[tp]
\begin{tabular}{  l  rr   rr   rr   rr    rr }
\toprule
   \multirow{2}{*}{System}  &\multicolumn{1}{c}{$R$}  &\multicolumn{1}{c}{$a$}   & \multicolumn{1}{c}{$P$} &  \multicolumn{1}{c}{$B_\star$} &  \multicolumn{1}{c}{$x$}& \multicolumn{1}{c}{$Q_{\rm abs}$}& \multicolumn{1}{c}{$t_{\rm ev,c}$} & \multicolumn{1}{c}{$t_{\rm ev,g}$} &\multicolumn{1}{c}{$t_{\rm ev,c}$}    & \multicolumn{1}{c}{$t_{\rm ev,g}$}  \\
  & [km] &[$R_\odot$]& [ms] &\multicolumn{1}{c}{[$10^{8}\,$G]}  &&  \multicolumn{1}{c}{\mbox{ }}&\multicolumn{2}{c}{[yr] $(\rho\sigma=10^6\,{\rm s^{-1}})$ } &\multicolumn{2}{c}{[yr] $(\rho\sigma=10^7\,{\rm s^{-1}})$}   \\

\midrule

       bodies  very  &    1.  &        1.  &   1. &  1.&    0.003      & $3.\times10^{-5}$ &   20.  &  ($2.\times10^{-6}$)&   0.6 &($7.\times10^{-8}$)\\ 
 
       near  1ms     &  $10^2$  &        1.  &   1. &  1.&    0.33       & $9.\times10^{-2}$ &    0.6 &   (0.007)    &6 &($0.007$) \\  
       pulsar        & $10^3$  &        1.  &   1. &  1.&    3.3        &   0.11        &    5.5  &   (0.6)    &54  &(6.)  \\ \midrule
       asteroid      &    1.  &        1.  &  10. &  1.& $3.\times10^{-4}$ & $3.\times10^{-7}$ & $2.\times10^{7}$&   (2.)&$6.\times10^{5}$&(0.06)\\  
       and 10\,ms PSR &   1.  &      200.  &  10. &  1.& $3.\times10^{-4}$ & $3.\times10^{-7}$ & $8.\times10^{11}$& ($9.\times10^{4}$)&$8.\times10^{11}$& $(9.\times 10^4)$\\ \midrule
       asteroid        + 1\,s PSR       &  1.&   1.  &   $10^3$ & $10^{4}$&$3.\times10^{-6}$  & $3.\times10^{-11}$ & $2.\times10^{11}$&   ($2.\times10^{4}$)&$2.4\times10^{10}$ & ($2.7\times10^{3}$) \\   \midrule
       asteroid      + PSR 1931+24
       &  1.& 0.14  & 813. & $10^{4}$&$4.\times10^{-6}$  & $5.\times10^{-11}$& $1.1 \times 10^{9}$&   (128) & $6.\times10^{9}$&(683)\\  \midrule
       PSR 1257+12a  &    1800. &      42. &   6. &$8.8$&    1.      &   0.06    & $5. \times 10^{5}$ &  ($1.7\times10^{5}$)& $4.\times10^{6}$&$1.7\times 10^6$\\
       PSR 1257+12b  &   $10^4$ &      77. &   6. &$8.8$&    5.5     &   0.04    & ($1.5\times10^{7}$)&  $1.6\times10^{8}$&$1.5\times10^{8}$&$1.6\times10^{9}$\\
       PSR 1257+12c  &   $10^4$ &      98. &   6. &$8.8$&    5.5     &   0.04    & ($2.3\times10^{7}$)&  $2.6\times10^{8}$&($2.3\times10^{8}$)&$2.6\times10^{9}$\\    

\bottomrule
\end{tabular}
\vspace{0.3cm}
\caption{Gravitational ($t_{\rm ev,g}$) and cohesive ($t_{\rm ev,c}$) evaporation timescales due to induction heating for various pulsar and companion parameters. We assume a mass density $\rho=3$ g cm$^{-3}$ and $\mu \sigma=10^6,10^7$ s$^{-1}$ for columns ($8-9$) and ($10-11$), respectively. Values of $Q_{\rm abs}$ were computed with {\tt Damie}. Timescale values in parentheses are only indicative as the other timescale dominates the evaporation regime.}
\label{table_temps_evaporation_3}
\end{table}
\end{landscape}

\subsection{Evaporation of small non-degenerate companions}\label{sec_lightcomp}

In the case of small objects like asteroids, Eq.~(\ref{eq:tevg}) fails because the dominant process for evaporating the body is no longer gravitational escape but heating, melting, and evaporation. { The cohesive energy can be expressed as 
\begin{equation}
E_{\rm c}=KM \sim 1.6\times 10^{44}\,{\rm erg}\,\frac{K}{K_{\rm iron}}\frac{M}{M_\odot}\ ,
\end{equation} 
with $K$ the cohesive factor that depends on the material \citep{Kittel_issp}. In particular, the cohesive factor of iron is $K_{\mathrm{iron}}=7.4\times 10^{10}$ erg/g (and the orders of magnitude are similar for rock).

Equating with $E_{\rm g}$,   the cohesive energy starts to dominate for a mass-to-size ratio of $M/R\lesssim 10^{18}\,$g\,cm$^{-1}$. For ordinary matter, this constraint can then be expressed in terms of companion size as
\begin{equation}
R\lesssim 5\times 10^8\,{\rm cm} \,\rho^{-1/2}\left(\frac{K}{K_{\rm iron}}\right)^{1/2}\ .
\end{equation}In this regime, the evaporation timescale can be calculated}
\begin{equation}
KM = Q_{\rm abs}(f/f_{\rm p})(R/2a)^2L_{\rm p} t_{\rm ev 3} 
\end{equation}
which leads to 
\begin{eqnarray} \label{eq:tevc}
\frac{t_{\rm ev,c}}{\rm yr} &=&2.6 \times 10^{-4}  \left(\frac{R}{\rm km}\right)  \left(\frac{\rho}{{\rm g\,cm}^{-3}}\right)  \left( \frac{K}{K_{\rm iron}}\right)   \left(\frac{a}{R_\odot}\right)^{2} \times \nonumber\\
& &\frac{f}{f_{\rm p}}\frac{1}{L_{\rm p,35}{Q_{\rm abs}}} \ .
\end{eqnarray}

For small objects, and when the pulsar energy is mostly in the form of Poynting flux, $Q_{\rm abs}$ is low and has a strong influence on the evaporation time.
The comparison of ${t_{\rm ev, g}}$ and ${t_{\rm ev, c}}$  shows, as expected, that when small bodies have melted and vaporized, the gravitational escape that follows is very fast.

Figure~\ref{fig:tev} shows the evaporation timescales of non-degenerate pulsar companions ($t_{\rm ev,g}$ and $t_{\rm ev,c}$, Eqs.~\ref{eq:tevg} and \ref{eq:tevc}), for three spin periods. 
For small companions, on the left-hand side of the white dotted line, the cohesive energy dominates over gravitational energy for evaporation. The green contours indicate evaporation timescales shorter than $\sim 10^6\,$yrs. A few typical examples are also given in Table~\ref{table_temps_evaporation_3} for  companions made of rock (columns $8-9$) and for a more metallic composition (columns $10-11$).

The table shows that small non-degenerate bodies at one solar radius from a 1\,ms pulsar evaporate in less than a few years. We note that for $P=1-100\,$ms and $N_{\rm r}\sim 100$, the fastest evaporation occurs for $R \sim 100\,$km,  and not for the smallest companions. We also note also the strong dependency of the evaporation time on the pulsar period $P$. An asteroid at the same distance of a 10\,ms pulsar can survive  a few $10^7$ years. At 1 AU (200\,$R_\odot$), it is definitively stable.

Figure~\ref{fig:tevNr} demonstrates that $N_{\rm r}$, namely the chemical composition of the companion, has a strong influence on the evaporation timescale. An asteroid ($R \sim 1$ km) made of rock ($N_{\rm r}\sim 100$ or $\mu \sigma=10^6$ s$^{-1}$) has an optimal lifetime. More resistive ($N_{\rm r}\sim 10$), or more metallic asteroids ($ N_{\rm r}\gtrsim 10^3$) evaporate over a shorter timescale. For $N_{\rm r}\lesssim 100$, the distance $a$ remains constant for small size ratios $x$ (black line in Fig.~\ref{fig:tevNr}).
This stems from the change in the dependency of $Q_{\rm abs}$ over $x$ (and thus over $R$), as shown in Fig.~\ref{fig_Q_abs_metalic} for small $x$ and low $N_{\rm r}$.

The comparison between columns ($8-9$) and ($10-11$) of Table~\ref{table_temps_evaporation_3} confirms that kilometer-size companions are more stable against evaporation when made of rock ($\mu \sigma=10^6$ s$^{-1}$) than for a more metallic refractive index $\mu \sigma=10^7$ s$^{-1}$. On the other hand,     the table and  Figure~\ref{fig:tev} show that for $R \ge 100$ km the evaporation timescale increases with the conductivity of the companion.

\cite{Mottez13} proposed that the intermittency of PSR 1931+24 can be explained by the existence of a stream of small bodies of kilometer or subkilometer sizes close to the pulsar ($a=0.14\,R_\odot$). We  calculate  that these asteroids would evaporate in about one Gyr. If the Mie theory were not taken into account, such asteroids would not survive a year. A billion years being a long timescale compared to the migration time in the pulsar wind magnetic field (10,000 years, \citealp{Mottez_2011_AWO}), the evaporation caused by induction does not invalidate the model of \cite{Mottez13}. 

The last lines in Table \ref{table_temps_evaporation_3} refers to planets discovered near PSR 1257+12 \citep{Wolszczan_1994}. Because of their size, the Mie theory still has an influence on the vaporization rate, with $Q_{\rm abs} < 0.06$.
 If it is made of rock, 1257+21a should evaporate in $10^5$ years. With a more metallic composition, it could last a few million years. The other two planets are stable for $10^8$ years, or more if they are metallic.

\begin{figure*}[th]
\begin{center}
\includegraphics[width=\textwidth]{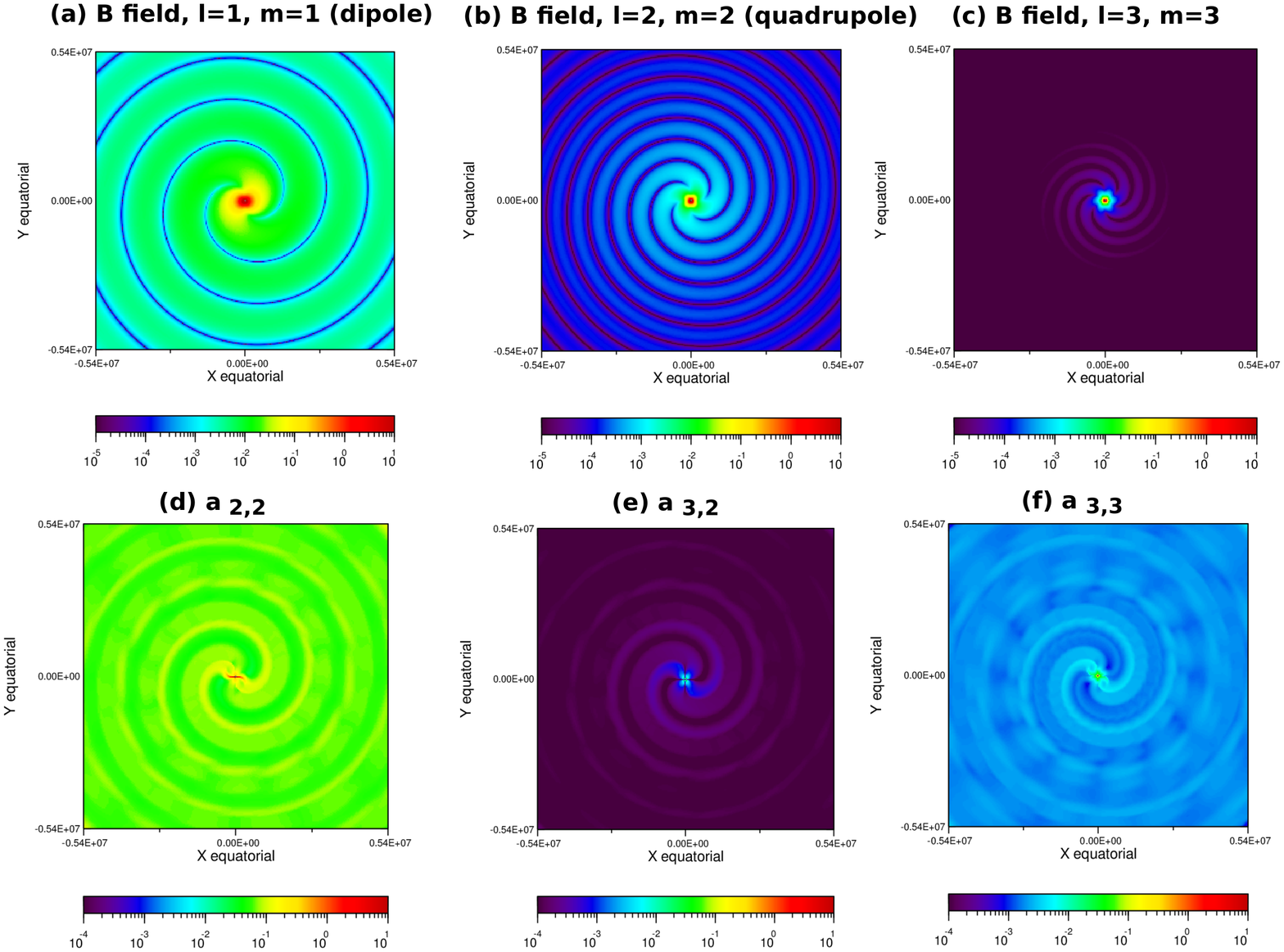}
\end{center}
\caption{Amplitude of the pulsar wind magnetic field for (a) a surface dipole field $l=1, m=1$, (b) quadrupole $l=2, m=2,$ and (c) $l=3, m=3$.
Panels (a) to (c) are plotted with the same color scale. On these panels, the spiral structure is the pulsar wave. Comparison ratio $a$, see Eq.~(\ref{def_a}) for modes (d) $l=2, m=2$, (e) $l=3, m=2,$ and (f) $l=3, m=3$. Panels (d) to (f) are plotted with the same color scale.}
\label{fig:multipole}
\end{figure*}

\section{Effect of multipole magnetic field components} \label{sec_multipole}
We have considered  the case of a star with a dipole magnetic field.
In terms of spherical harmonics analysis, it has longitudinal and azimuthal components $l=1$, $m=0$, and $m=1$. 
Only the $m=1$ component participates in the pulsar wave and produces a monochromatic frequency spectrum,  
but we know that neutron star magnetic fields can have multipole components $l>1$. 
The components with azimuthal numbers $m>1$ induce shorter wavelengths $cP/m$, i.e., $2 \pi R_{\rm L}/m$. In all the above inductive heating rates,  a multipole term of azimuthal number $m$ can be accounted for by replacing the ratio of the size of the body to the incident wavelength $x$ by $x_m=m x$.
If they had the same amplitude,  multipole $m>1$ terms would contribute more efficiently to the companion heating than the $m=1$ wave for 
small bodies where {\kumiko $x \sim 1$}. 
Therefore, it is important to evaluate the contribution of multipole terms of the neutron star magnetic field to the pulsar wind. 

\citet{Mottez_2015c} have computed the vacuum solution of the electromagnetic field surrounding a pulsar. The solutions have a complex structure at distances less than the light cylinder radius with an amplitude that decreases faster for higher values of $l$. At greater distances, it has the characteristic Parker spiral wave structure dominated by the azimuthal component $B_\phi \varpropto r^{-1}$. Beyond the light cylinder, the ratio of amplitudes of the various modes does not vary much on average. To illustrate this, we have plotted in Fig. \ref{fig:multipole}  a few  vacuum field solutions corresponding to the same maximum amplitude $B_0$ (on the star surface).  The amplitude $B$ of the magnetic field in the equatorial plane is plotted  up to distances $r=11 R_L$ for the dipole solution (panel a), for the quadrupole solution $l=2 m=2$ (panel b)\footnote{A pulsar of period $P_4$ with a pure quadrupole field might be confused, from an observational point of view, with a pulsar with a dipole field and a period $P_2=2 P_4$.}, and for $l=3, m=3$ (panel c). For comparison of magnetic amplitudes, we define the ratio 
\begin{eqnarray} \label{def_a}
 a_{lm}(R,\phi)=\max(B_{lm}(r>R,\phi,\theta=0))/ \\\max(B_{\rm dipole}(r>R,\phi,\theta=0))
,\end{eqnarray}
where the maximum is computed for any radius larger than $R$. This definition is intended to avoid a divergent ratio when the two functions reach their minimum 
for different values of $R$ that would not be pertinent to what we want to compare.
The ratios $a_{2,2}, a_{3,2}$, and $a_{3,3}$ are plotted respectively in panels (d), (e), and (f).
We can see from $a_{2,2}$ that beyond a distance $2 R_L$, the contributions of $l=1, m=1$ (dipole) and $l=2, m=2$ modes with the same maximum magnetic field on the star's surface are comparable, although larger with the dipole. For larger values of $l$, the dipole term is dominant by a factor exceeding $10^2$. 
This means that the contribution of a large amplitude quadrupole term to the heating of a pulsar companion cannot be neglected, implying that companions of radius $R/2$ twice as small can absorb an energy comparable to that of a $R$-sized body in the dipole case. Higher multipole terms should not have an important influence in our example. 

We note that the above computation does not take into account the contribution of the plasma to the evolution of the wave amplitude into the wind; it should  be considered  a first-order approximation.

\section{Conclusion and discussion}
In this paper, we have assessed the efficiency of the inductive heating of pulsar companions caused by the pulsar Poynting-flux electromagnetic wave. Inductive heating is generally considered to be the dominant heating process, for instance when the evaporation of black widow companions is considered \citep{vandenHeuvel88}. It is commonly assumed that the whole flux carried by the pulsar wave is absorbed by the companion. Taking into account the Mie theory, we have shown that this assumption fails for objects with radius $R$ smaller than the pulsar wavelength $c P$, $P$ being the pulsar spin period. The rate of absorption of the wave energy decreases by many orders of magnitude for small objects like asteroids or planetesimals. This is especially true with standard $P=1\,$s pulsars because of their long wavelength. One consequence is that asteroids, considered to be objects that should quickly evaporate, do not always do so even at  very close distances ($1 R_\odot$) of a $1$\,s-period pulsar. 
 
The conclusion regarding kilometer-sized asteroids at short distances from  pulsars cannot be generalized to any kind of pulsar. For instance, bodies orbiting a $P=1$ ms pulsar at $1 R_\odot$ would not last a year.

The behavior of small objects orbiting a pulsar is important because it can  explain the intermittency of their radio-emissions. For instance, \cite{Cordes08} proposed a model where  small rocks (meter-sized) falling on a pulsar cause intermittency during the few seconds in which they evaporate. For metric objects, $Q_{\rm abs}$ derived from the Mie theory is so low that these rocks should not evaporate. 

The survival of small bodies around pulsars is a pre-requisite of the theory developed by \cite{Mottez14} to explain the origin of FRBs. In that model, asteroids immersed in the pulsar relativistic wind produce collimated magnetic wakes, that can be observed at cosmological distances and lead to radio bursts similar to the observed signals. An asteroid belt could also be invoked to account for the repeating FRB reported by \cite{Spitler16} (see also \citealp{Dai16}).

We have assumed in this paper that the wind was Poynting-flux dominated at the location of the companion. For objects made of ordinary matter, the consistency of our evaporation timescales compared to the observations seems to validate a posteriori this hypothesis. If a non-negligible fraction of the wind energy were kinetic, a heating process by high-energy particle irradiation would have to be considered. This process would evaporate most close-by companions, whatever their size, as has been calculated and observed for black widow and redback systems (e.g., \citealp{Phinney88,Kluzniak88, Arons93}). In this case, the energy absorption coefficient $Q_{\rm kin}$ can be taken as $\sim 1$ at first-order approximation. A more thorough evaluation of $Q_{\rm kin}$ would require  calculating the cascading interactions of particles, which is beyond the scope of this study. Models considering simultaneously inductive heating, sputtering by the wind, and blackbody X-rays will be developed in forthcoming papers. 
From Eqs.~(\ref{eq:Qabs_decomp}) and (\ref{eq:tevc}), one can then infer that in order to avoid heating by particles, the fraction of kinetic energy in the wind has to be lower than $Q_{\rm em}$, the absorption coefficient of the Poynting flux given by the Mie Theory (denoted $Q_{\rm abs}$ by abuse of notation throughout this paper, see values in Fig.~\ref{fig_Q_abs_metalic} and Table~\ref{table_temps_evaporation_3}). This implies a magnetization of $\sigma_B>1/Q_{\rm em}-1$. As $Q_{\rm em} \ll 1$ over a large parameter space, as is demonstrated in this paper, a very low fraction of kinetic energy in the wind and thus a very high magnetization is required for the survival of objects close to the pulsar. 

We note that we have also argued in this work that, because of their high refractive index to the pulsar wave, the observed ablation of white dwarfs and M-stars in back widow and redback systems can only be explained if the absorbed energy is principally kinetic.

\section*{Acknowledgement}
This work was nurtured by Jean Heyvaerts' notes on the Mie theory and by the inspiring discussions we had with him. His great insight and input were very much missed during the completion of the project. KK thanks S. Phinney for fruitful discussions that led to her involvement in this project. This work was supported by the Programme National des Hautes Energies. 
KK acknowledges financial support from the PER-SU fellowship at Sorbonne Universit\'es, from the Labex ILP (reference ANR-10-LABX-63, ANR-11-IDEX-0004-02), as well as from the NSF grant NSF PHY-1412261 and the NASA grant 11-APRA-0066 at the University of Chicago, and the grant NSF PHY-1125897 at the Kavli Institute for Cosmological Physics.

\bibliographystyle{aa} 

\bibliography{KO}

\begin{thebibliography}{42}
\expandafter\ifx\csname natexlab\endcsname\relax\def\natexlab#1{#1}\fi

\bibitem[{{Alpar} {et~al.}(1982){Alpar}, {Cheng}, {Ruderman}, \&
  {Shaham}}]{Alpar82}
{Alpar}, M.~A., {Cheng}, A.~F., {Ruderman}, M.~A., \& {Shaham}, J. 1982,
  Nature, 300, 728

\bibitem[{{Arons} \& {Tavani}(1993)}]{Arons93}
{Arons}, J. \& {Tavani}, M. 1993, ApJ, 403, 249

\bibitem[{{Bailes} {et~al.}(2011){Bailes}, {Bates}, {Bhalerao}, {Bhat},
  {Burgay}, {Burke-Spolaor}, {D'Amico}, {Johnston}, {Keith}, {Kramer},
  {Kulkarni}, {Levin}, {Lyne}, {Milia}, {Possenti}, {Spitler}, {Stappers}, \&
  {van Straten}}]{Bailes11}
{Bailes}, M., {Bates}, S.~D., {Bhalerao}, V., {et~al.} 2011, Science, 333, 1717

\bibitem[{{Bonazzola} {et~al.}(2015){Bonazzola}, {Mottez}, \&
  {Heyvaerts}}]{Mottez_2015c}
{Bonazzola}, S., {Mottez}, F., \& {Heyvaerts}, J. 2015, \aap, 573, A51

\bibitem[{{Born} \& {Wolf}(1980)}]{Born_1980}
{Born}, M. \& {Wolf}, E. 1980, {Principles of Optics Electromagnetic Theory of
  Propagation, Interference and Diffraction of Light}, chap. Optics of metals

\bibitem[{{Callanan} {et~al.}(1995){Callanan}, {van Paradijs}, \&
  {Rengelink}}]{Callanan95}
{Callanan}, P.~J., {van Paradijs}, J., \& {Rengelink}, R. 1995, ApJ, 439, 928

\bibitem[{{Canuto} \& {Solinger}(1970)}]{Canuto70}
{Canuto}, V. \& {Solinger}, A.~B. 1970, ApJ Lett., 6, 141

\bibitem[{{Champion} {et~al.}(2016){Champion}, {Petroff}, {Kramer}, {Keith},
  {Bailes}, {Barr}, {Bates}, {Bhat}, {Burgay}, {Burke-Spolaor}, {Flynn},
  {Jameson}, {Johnston}, {Ng}, {Levin}, {Possenti}, {Stappers}, {van Straten},
  {Thornton}, {Tiburzi}, \& {Lyne}}]{Champion16}
{Champion}, D.~J., {Petroff}, E., {Kramer}, M., {et~al.} 2016, MNRAS

\bibitem[{{Cordes} \& {Shannon}(2008)}]{Cordes08}
{Cordes}, J.~M. \& {Shannon}, R.~M. 2008, ApJ, 682, 1152

\bibitem[{{Dai} {et~al.}(2016){Dai}, {Wang}, {Wu}, \& {Huang}}]{Dai16}
{Dai}, Z.~G., {Wang}, J.~S., {Wu}, X.~F., \& {Huang}, Y.~F. 2016, ArXiv
  e-prints: 1603.08207

\bibitem[{{Deneva} {et~al.}(2009){Deneva}, {Cordes}, \& {Lazio}}]{Deneva09}
{Deneva}, J.~S., {Cordes}, J.~M., \& {Lazio}, T.~J.~W. 2009, ApJ Lett., 702,
  L177

\bibitem[{{Deutsch}(1955)}]{Deutsch_1955}
{Deutsch}, A.~J. 1955, Annales d'Astrophysique, 18, 1

\bibitem[{{Gough}(1989)}]{Gough_1989}
{Gough}, D.~I. 1989, Reviews of Geophysics, 27, 141

\bibitem[{{Haensel} {et~al.}(1999){Haensel}, {Lasota}, \& {Zdunik}}]{Haensel99}
{Haensel}, P., {Lasota}, J.~P., \& {Zdunik}, J.~L. 1999, A\&A, 344, 151

\bibitem[{{Huang} \& {Geng}(2014)}]{Huang14}
{Huang}, Y.~F. \& {Geng}, J.~J. 2014, ApJ Lett., 782, L20

\bibitem[{{Khechinashvili} {et~al.}(2000){Khechinashvili}, {Melikidze}, \&
  {Gil}}]{Khechinashvili00}
{Khechinashvili}, D.~G., {Melikidze}, G.~I., \& {Gil}, J.~A. 2000, ApJ, 541,
  335

\bibitem[{{Kirk} {et~al.}(2009){Kirk}, {Lyubarsky}, \& {Petri}}]{Kirk09}
{Kirk}, J.~G., {Lyubarsky}, Y., \& {Petri}, J. 2009, in Astrophysics and Space
  Science Library, Vol. 357, Astrophysics and Space Science Library, ed.
  W.~{Becker}, 421

\bibitem[{Kittel(1998)}]{Kittel_issp}
Kittel, C. 1998, Physique de l'{\'e}tat solide, {\'e}dition : 7{\`e}me
  {\'e}dition edn. (Paris: Dunod)

\bibitem[{{Kluzniak} {et~al.}(1988){Kluzniak}, {Ruderman}, {Shaham}, \&
  {Tavani}}]{Kluzniak88}
{Kluzniak}, W., {Ruderman}, M., {Shaham}, J., \& {Tavani}, M. 1988, Nature,
  334, 225

\bibitem[{{Lentz}(1976)}]{Lentz_1976}
{Lentz}, W.~J. 1976, Applied Optics, 15, 668

\bibitem[{{Manchester} {et~al.}(2005){Manchester}, {Hobbs}, {Teoh}, \&
  {Hobbs}}]{Manchester05}
{Manchester}, R.~N., {Hobbs}, G.~B., {Teoh}, A., \& {Hobbs}, M. 2005, VizieR
  Online Data Catalog, 7245, 0

\bibitem[{{Mottez} {et~al.}(2013{\natexlab{a}}){Mottez}, {Bonazzola}, \&
  {Heyvaerts}}]{Mottez13_a}
{Mottez}, F., {Bonazzola}, S., \& {Heyvaerts}, J. 2013{\natexlab{a}}, \aap,
  555, A125

\bibitem[{{Mottez} {et~al.}(2013{\natexlab{b}}){Mottez}, {Bonazzola}, \&
  {Heyvaerts}}]{Mottez13}
{Mottez}, F., {Bonazzola}, S., \& {Heyvaerts}, J. 2013{\natexlab{b}}, A\&A,
  555, A126

\bibitem[{{Mottez} \& {Heyvaerts}(2011)}]{Mottez_2011_AWO}
{Mottez}, F. \& {Heyvaerts}, J. 2011, Astronomy and Astrophysics, 532, A22+

\bibitem[{{Mottez} \& {Zarka}(2014)}]{Mottez14}
{Mottez}, F. \& {Zarka}, P. 2014, A\&A, 569, A86

\bibitem[{{Phinney} {et~al.}(1988){Phinney}, {Evans}, {Blandford}, \&
  {Kulkarni}}]{Phinney88}
{Phinney}, E.~S., {Evans}, C.~R., {Blandford}, R.~D., \& {Kulkarni}, S.~R.
  1988, Nature, 333, 832

\bibitem[{{Ransom} {et~al.}(2014){Ransom}, {Stairs}, {Archibald}, {Hessels},
  {Kaplan}, {van Kerkwijk}, {Boyles}, {Deller}, {Chatterjee},
  {Schechtman-Rook}, {Berndsen}, {Lynch}, {Lorimer}, {Karako-Argaman}, {Kaspi},
  {Kondratiev}, {McLaughlin}, {van Leeuwen}, {Rosen}, {Roberts}, \&
  {Stovall}}]{Ransom14}
{Ransom}, S.~M., {Stairs}, I.~H., {Archibald}, A.~M., {et~al.} 2014, Nature,
  505, 520

\bibitem[{{Reynolds} {et~al.}(2007){Reynolds}, {Callanan}, {Fruchter},
  {Torres}, {Beer}, \& {Gibbons}}]{Reynolds07}
{Reynolds}, M.~T., {Callanan}, P.~J., {Fruchter}, A.~S., {et~al.} 2007, MNRAS,
  379, 1117

\bibitem[{{Roberts}(2013)}]{Roberts13}
{Roberts}, M.~S.~E. 2013, in IAU Symposium, Vol. 291, IAU Symposium, ed.
  J.~{van Leeuwen}, 127--132

\bibitem[{{Ruderman} {et~al.}(1989){Ruderman}, {Shaham}, \&
  {Tavani}}]{Ruderman89}
{Ruderman}, M., {Shaham}, J., \& {Tavani}, M. 1989, ApJ, 336, 507

\bibitem[{{Savonije}(1987)}]{Savonije87}
{Savonije}, G.~J. 1987, Nature, 325, 416

\bibitem[{{Shannon} {et~al.}(2013){Shannon}, {Cordes}, {Metcalfe}, {Lazio},
  {Cognard}, {Desvignes}, {Janssen}, {Jessner}, {Kramer}, {Lazaridis},
  {Purver}, {Stappers}, \& {Theureau}}]{Shannon13}
{Shannon}, R.~M., {Cordes}, J.~M., {Metcalfe}, T.~S., {et~al.} 2013, ApJ, 766,
  5

\bibitem[{{Shapiro} \& {Teukolsky}(1983)}]{Shapiro83}
{Shapiro}, S.~L. \& {Teukolsky}, S.~A. 1983, {Black holes, white dwarfs, and
  neutron stars: The physics of compact objects} (John Wiley and Son. Inc.)

\bibitem[{{Sigurdsson} {et~al.}(2003){Sigurdsson}, {Richer}, {Hansen},
  {Stairs}, \& {Thorsett}}]{Sigurdsson03}
{Sigurdsson}, S., {Richer}, H.~B., {Hansen}, B.~M., {Stairs}, I.~H., \&
  {Thorsett}, S.~E. 2003, Science, 301, 193

\bibitem[{{Spitler} {et~al.}(2016){Spitler}, {Scholz}, {Hessels}, {Bogdanov},
  {Brazier}, {Camilo}, {Chatterjee}, {Cordes}, {Crawford}, {Deneva}, {Ferdman},
  {Freire}, {Kaspi}, {Lazarus}, {Lynch}, {Madsen}, {McLaughlin}, {Patel},
  {Ransom}, {Seymour}, {Stairs}, {Stappers}, {van Leeuwen}, \&
  {Zhu}}]{Spitler16}
{Spitler}, L.~G., {Scholz}, P., {Hessels}, J.~W.~T., {et~al.} 2016, Nature,
  531, 202

\bibitem[{{Stappers} {et~al.}(2003){Stappers}, {Gaensler}, {Kaspi}, {van der
  Klis}, \& {Lewin}}]{Stappers03}
{Stappers}, B.~W., {Gaensler}, B.~M., {Kaspi}, V.~M., {van der Klis}, M., \&
  {Lewin}, W.~H.~G. 2003, Science, 299, 1372

\bibitem[{{Thorsett} {et~al.}(1993){Thorsett}, {Arzoumanian}, \&
  {Taylor}}]{Thorsett93}
{Thorsett}, S.~E., {Arzoumanian}, Z., \& {Taylor}, J.~H. 1993, ApJ Lett., 412,
  L33

\bibitem[{{van de Hulst}(1981)}]{Hulst_81}
{van de Hulst}, H.~C. 1981, {Light scattering by small particles}

\bibitem[{{van den Heuvel} \& {van Paradijs}(1988)}]{vandenHeuvel88}
{van den Heuvel}, E.~P.~J. \& {van Paradijs}, J. 1988, Nature, 334, 227

\bibitem[{{van Kerkwijk} {et~al.}(2011){van Kerkwijk}, {Breton}, \&
  {Kulkarni}}]{vanKerkwijk11}
{van Kerkwijk}, M.~H., {Breton}, R.~P., \& {Kulkarni}, S.~R. 2011, ApJ, 728, 95

\bibitem[{{Wolszczan}(1994)}]{Wolszczan_1994}
{Wolszczan}, A. 1994, Science, 264, 538

\bibitem[{{Wolszczan} \& {Frail}(1992)}]{Wolszczan92}
{Wolszczan}, A. \& {Frail}, D.~A. 1992, Nature, 355, 145

\end{thebibliography}

\end{document}